\documentclass[10pt,prl,aps,twocolumn,tightenlines,superscriptaddress,nofootinbib,notitlepage,preprintnumbers]{revtex4-1}
\pdfoutput=1
\usepackage{epsfig,amsfonts,mathrsfs,amsmath,amssymb,graphicx,color}

\begin{document}
\title{A Cosmic Selection Rule for Glueball Dark Matter Relic Density}
\author{Amarjit Soni}
\email{soni@bnl.gov}
\affiliation{Physics Department, Brookhaven National Laboratory, Upton, NY 11973, USA}
\author{Huangyu Xiao}
\email{xiaohy13@mails.tsinghua.edu.cn}
\affiliation{Department of Physics, Tsinghua University, Beijing, China}
\author{Yue Zhang}
\email{yuezhang@northwestern.edu}
\affiliation{Department of Physics and Astronomy, Northwestern University, Evanston, IL 60208, USA}

\begin{abstract}
We point out a unique mechanism to produce the relic abundance for glueball dark matter from a gauged $SU(N)_d$ hidden sector which is bridged to the standard model sector through heavy vectorlike quarks colored under gauge interactions from both sides. A necessary ingredient of our assumption is that the vectorlike quarks, produced either thermally or non-thermally, are abundant enough to dominate the universe for some time in the early universe. They later undergo dark color confinement and form unstable vectorlike-quarkonium states which annihilate decay and reheat the visible and dark sectors. The ratio of entropy dumped into two sectors and the final energy budget in the dark glueballs is only determined by low energy parameters, including the intrinsic scale of the dark $SU(N)_d$, $\Lambda_d$, and number of dark colors, $N_d$, but depend weakly on parameters in the ultraviolet such as the vectorlike quark mass or the initial condition. We call this a cosmic selection rule for the glueball dark matter relic density.
\end{abstract}
\preprint{NUHEP-TH/17-03}

\maketitle
\pagestyle{plain}

\section{Introduction}

A candidate for dark matter (DM) in the universe is the lightest glueball from a dark sector with pure $SU(N)_d$ gauge symmetry. Such a setup offers a natural explanation of the DM mass scale through dimensional transmutation. It is also extremely simple; the glueball DM properties only depend on two parameters, the $SU(N)_d$ intrinsic scale, $\Lambda_d$, and the number of dark colors, $N_d$~\cite{Soni:2016gzf}. Moreover, the interactions of glueball DM could offer a series of phenomena allowing it to be examined in cosmology and astrophysics. They include self-interacting and warm DM scenarios~\cite{Carlson:1992fn, Boddy:2014yra, Buen-Abad:2015ova, Soni:2016gzf}, indirect detection signatures when the glueball DM decays into the standard model (SM) particles~\cite{Juknevich:2009ji}, and even the formation of compact stars as the source of gravitational waves~\cite{Soni:2016yes, daRocha:2017cxu}. For the above reasons, such a DM candidate has been often revisited in recent years.

Among the many aspects of glueball DM, its production mechanism in the early universe has been less explored. Due to gauge invariance, higher dimensional operators are needed for such a dark sector to communicate with the SM sector. This implies that the abundance of glueball DM, if ever related to that of the SM sector, would be highly sensitive to the detailed high scale physics, or the initial conditions~\cite{Faraggi:2000pv, Yamanaka:2014pva, Forestell:2016qhc, Halverson:2016nfq, bobby}. One could resort to ultraviolet (UV) complete models by introducing new particles and portals between the dark and SM sectors. However, if the universe had been hot enough and a thermal equilibrium established between the two sectors, glueball DM that is akin to us, if much heavier than neutrinos, would be easily overproduced because they are in the form of dark gluons and relativistic when decoupling from the SM sector. 

In this paper, we point out a novel glueball DM production mechanism. We first consider the case where the dark and SM sectors thermalize in the early universe through a pair of vectorlike quarks colored under both the dark $SU(N)_d$ and the strong interaction $SU(3)_c$. The key to produce correct relic abundance of the glueball DM is to have the vectorlike quarks first freeze out and later become the dominant species in the energy density of the universe. Such a matter dominated phase will end when the vectorlike quarks in the universe is diluted enough and dark confinement happens binding them into composite states neutral under $SU(N)_d$. For large enough $N_d$, the most formed composite states are the vectorlike-quarkonium (which we will call $Q$-onium) states made of $Q$ and $\bar Q$. They are unstable and will quickly decay away, via $Q\bar Q$ annihilation, into quarks, gluons, and dark gluons. This will produce entropy and reheat the two sectors according to the $Q$-onium branching ratios, afterwards the standard cosmology in the SM sector could begin. Interestingly, we find that the resulting glueball DM relic abundance is mainly determined by low energy parameters including the intrinsic scale $\Lambda_d$ and number of dark colors $N_d$, but depends very weakly on detailed high scale physics such as the vectorlike quark mass. More generally, we also argue that, as long as the vectorlike quark dominated universe occurs, the very same conclusion would still hold for non-thermal initial conditions.

\section{Model}

We consider a dark sector, which at low energy scale is made of a pure $SU(N)_d$ gauge symmetry, with an intrinsic scale $\Lambda_d$ where the dark gauge coupling goes strong. The DM candidate within this setup is the lightest dark glueball state, $\phi$, which is assumed to be a scalar particle based on existing lattice studies~\cite{Cornwall:1982zn, Morningstar:1999rf, Lucini:2010nv, Lucini:2001ej}. If this is a completely isolated dark sector, $\phi$ could serve as the DM candidate. With a mass below $10^7$\,GeV, it can be cosmologically stable against decaying into two gravitons~\cite{Soni:2016gzf}. 

Because of gauge invariance, it is possible for such a dark sector to talk to the SM particles through higher dimensional operators, which could be generated by integrating out the high scale physics that bridges the two sectors. 
In this work, our goal is to establish a theoretical connection between the relic abundance of such a glueball DM, and that of the SM sector, which is insensitive to the cutoff scale of any higher dimensional operators.
To this end, we resort to a UV complete model by introducing vectorlike quarks $Q$, $\bar Q$ which play the role as bridge particles. Their quantum numbers under the $SU(3)_c\times SU(2)_L\times U(1)_Y\times SU(N)_d$ gauge groups are
\begin{equation}\label{quantumnumbers}
Q \in (3, 2, 1/6, N_d), \ \ \ \bar Q \in (\bar 3, \bar 2, -1/6, \bar N_d) \ .
\end{equation}
With $Q, \bar Q$ and their gauge interactions, it is possible to build a thermal contact between the two sectors in the early universe, when the temperature is high enough.

The presence of the vectorlike quarks opens a new decay channel of the glueball DM $\phi$ into two photons at loop level. The decay rate is~\cite{Juknevich:2009ji}
\begin{equation}
\Gamma_{\phi \to \gamma\gamma} = \frac{\alpha_d^2 \alpha^2 (N_d^2-1) m_\phi^2 F_\phi^2}{2^{9} 3^4 \pi m_Q^8} \ .
\end{equation}
The factor $F_\phi \equiv \langle0| \frac{1}{2} G^{a \mu\nu}_d G^a_{d \mu\nu}| \phi \rangle$ and $G_d$ is the field strength of the dark gluon field. Through naive dimensional analysis and large $N_d$ power counting, $F_\phi \sim N_d m_\phi^3$.
Hereafter, we will focus dark glueball mass below GeV scale, thus the hadronic decay channels are much less important.
The dark gauge coupling $\alpha_d=g_d^2/(4\pi)$ is evaluated at the mass scale $m_Q$,
\begin{equation}\label{alphaD}
\alpha_d(m_Q) =\frac{6\pi}{11 N_d \log(m_Q^2/\Lambda_d^2)} \ .
\end{equation}
Requiring the above radiative decay lifetime to satisfy the present constraints on gamma-ray or X-ray injection from DM decay $\tau_{\phi \to \gamma\gamma} \gtrsim 10^{28}\,$sec, 
we find a lower bound on the mass of $Q$,
\begin{equation}
m_Q \gtrsim 340\,{\rm GeV} \times \left( \frac{m_\phi}{1\,{\rm MeV}} \right)^\frac{9}{8} \left( \frac{N_d}{10^2} \right)^\frac{1}{4} \ ,
\end{equation}
where we have made the approximation that the logarithmic factor in $\alpha_d$ is of order 1.

\section{Cosmology}

For the early universe cosmology, we choose to first describe a thermal history. We will generalize the picture to be discussed to non-thermal initial conditions by the end of the next section. 

To begin with, we consider the temperature of the universe high enough and the SM sector, dark sector and the vectorlike quarks are all in thermal equilibrium with each other. This can be achieved through the annihilation processes $Q\bar Q \leftrightarrow gg,\, q\bar q,\, g' g'$, where $g\, (q)$ is the gluon (quark) in the SM sector and $g'$ is the  gluon in the dark $SU(N)_d$ sector. 
The condition for keeping all the particles in thermal equilibrium is, 
$\Gamma_{Q\bar Q\to f} = n_{Q} \left\langle (\sigma v_{\rm rel})_{Q\bar Q\to f} \right\rangle \gtrsim H$,
for all channels $f=\{gg,\, q\bar q,\, g' g'\}$, where $\langle\rangle$ means thermal averaging of the cross section times relative velocity. For $T>m_Q$, we have
$\Gamma_{Q\bar Q\to f} \propto T$ and the Hubble parameter $H \propto T^2$. Therefore, to ensure thermalization happens, it is sufficient to require the above condition to be satisfied around $T\sim m_Q$.

As the temperature falls below $m_Q$, the vectorlike quarks will freeze out by annihilating into both sectors.
The relevant $S$-wave cross sections, when $T\ll m_Q$, take the forms
\begin{eqnarray}\label{eq5}
\begin{split}
&(\sigma v_{\rm rel})_{Q\bar Q\to g g} = \frac{7\pi \alpha_S^2}{54 N_d m_Q^2} \ , \ \
(\sigma v_{\rm rel})_{Q\bar Q\to q\bar q} = 6\!\times\!\frac{\pi \alpha_S^2}{9 N_d m_Q^2}, \\
&(\sigma v_{\rm rel})_{Q\bar Q\to g' g'} = \frac{(N_d^2-1)(N_d^2-2) \pi \alpha_d^2}{48 N_d^3 m_Q^2} \ ,\\
&(\sigma v_{\rm rel})_{Q\bar Q\to g g'} = \frac{(N_d^2-1) 2\pi \alpha_S\alpha_d}{9 N_d^2 m_Q^2} \ ,
\end{split}
\end{eqnarray}
where both $\alpha_S$ and $\alpha_d$ are to be evaluated at the scale $m_Q$. It is worth noting that for $m_Q\gg \Lambda_d$, $\alpha_d$ goes as $\sim N_d^{-1}$ based on the RG equation, Eq.~(\ref{alphaD}). As a result, all the above cross sections are proportional to $N_d^{-1}$ in the large $N_d$ limit, and the annihilation processes into both SM and dark sectors will freeze out around the same epoch.
We have neglected the Sommerfeld enhancement effect which only modifies the above cross sections by an order one factor during the freeze out of $Q$~\cite{ElHedri:2016onc}.
We have also neglected the $Q\bar Q$ annihilation into two SM electroweak gauge bosons which are subdominant to the two gluon final state channel because of the smaller gauge couplings.
The Boltzmann equation governing the freeze out of $Q, \bar Q$ can be written as
\begin{eqnarray}
\frac{H}{s}\frac{d Y_{Q}}{d\ln z} =\sum_{f} \left\langle (\sigma v_{\rm rel})_{Q\bar Q\to f} \right\rangle \left[ Y_Q^2 - \left(Y_Q^{eq} \right)^2 \right] \ ,
\end{eqnarray}
where $z=m_Q/T$, $s$ is the entropy density of the universe receiving contribution from both sectors, the yield is defined as $Y_Q=n_Q/s$, and $n_Q$ is the number density of $Q$ particles. We assume no CP violation thus $Y_{\bar Q}=Y_Q$. At $T\gtrsim m_Q$, the number densities of $Q$ and $\bar Q$, which are still in thermal equilibrium, take the form
\begin{eqnarray}
n_Q = n_{\bar Q} =  12 N_d \times \frac{m_Q^2 T}{2\pi^2} K_2\left( \frac{m_Q}{T} \right) \ ,
\end{eqnarray}
where $K_2$ is the modified Bessel function of the second kind.
They serve as the initial conditions for solving the above Boltzmann equation.

Right after freeze out, the number of $Q, \bar Q$ are Boltzmann suppressed compared to the radiation species in the SM and dark sectors. Around this time, the two sectors share the same temperature $\mathcal{T}_f=T_f$ (hereafter we use $\mathcal{T}$ $(T)$ to denote the dark (SM) sector temperature). Their energy densities are
\begin{eqnarray}\label{rho123}
\begin{split}
&\rho_{Q+\bar Q} = 2 m_Q Y_Q(T_f) s(T_f) \ , \\
&\rho_{\rm SM} = \frac{\pi^2}{30} g_*^{\rm SM} T_f^4 \ , \ \ \
\rho_{\rm d} = \frac{\pi^2}{30} g_*^{\rm d} T_f^4 \ ,
\end{split}
\end{eqnarray}
where $g_*^{\rm SM}=106.75$ and $g_*^{\rm d}=2(N_d^2-1)$. As a convention, we define the universe expansion parameter corresponding to this temperature ($\mathcal{T}_f=T_f$), $a_{f}=1$.
Afterwards, the interactions between the two sectors can only go through off-shell $Q, \bar Q$ and quickly fall out of thermal equilibrium, so the two sectors will evolve with their own temperatures. 
At temperatures well below $m_Q$, interactions like $g Q\to g Q$, $g' Q\to g' Q$ are elastic, {\it i.e.}, they do not cause the exchange of heat among the three species, and the time dependence of energy densities in Eq.~(\ref{rho123}) still holds.

Next, we are interested in the fate of the vectorlike quarks $Q, \bar Q$ particles. After freeze out, they are already non-relativistic and their energy density $\rho_{Q+\bar Q}$ redshifts as $a^{-3}$, where $a$ is the expansion parameter of the universe. In contrast, $\rho_{\rm SM}$ and $\rho_{\rm d}$ redshifts faster, as $a^{-4}$. As a result, it is possible for $\rho_{Q+\bar Q}$ to come into dominance of the universe at a later stage. With our setup, the vectorlike quarks are stable particles. The only way to deplete them from the universe is to have $Q$ and $\bar Q$ find each other to annihilate. Well after the thermal freeze out, this could happen only because they are colored under the dark $SU(N)_d$ gauge group. Once their number densities fall below $\Lambda_d^3$, the dark confinement will take place and the $Q$ and $\bar Q$ particles will bind into color singlet states under $SU(N)_d$.~\footnote{There is also the QCD phase transition and $SU(3)_c$ color confinement which could happen before the dark sector one, but because there are many more SM quarks than the heavy vectorlike quarks in the universe, it is much more likely for $Q$ (or $\bar Q$) to find a light quark (or antiquark) than finding each other to become $SU(3)_c$ color neutral. Moreover, because $\Lambda_{\rm QCD}\ll m_Q$, the composite states made of $Q\bar q$ have approximately equal mass to $Q$. They are still colored under the dark $SU(N)_d$ and stable, and a dark sector phase transition is still necessary to eventually annihilate $Q$, $\bar Q$ away.} We call the expansion parameter corresponding to this event, $a_{c}$. It could be solved using
\begin{eqnarray}
a_c = \frac{1}{\Lambda_d} \left(2Y_Q(T_f) s(T_f) \rule{0mm}{3.6mm}\right)^\frac{1}{3} \ .
\end{eqnarray}
The corresponding dark sector temperature is $\mathcal{T}_c = T_f/a_c$, while the SM may have experienced a change in the degrees of freedom after freeze out, and its temperature $T_c$ could be obtained by solving the entropy conservation $g_{*S}^{\rm SM}(T_f) T_f^3 = g_{*S}^{\rm SM}(T_c) T_c^3 a_c^3$.

\begin{figure}[t]
\centerline{\includegraphics[width=0.95\columnwidth]{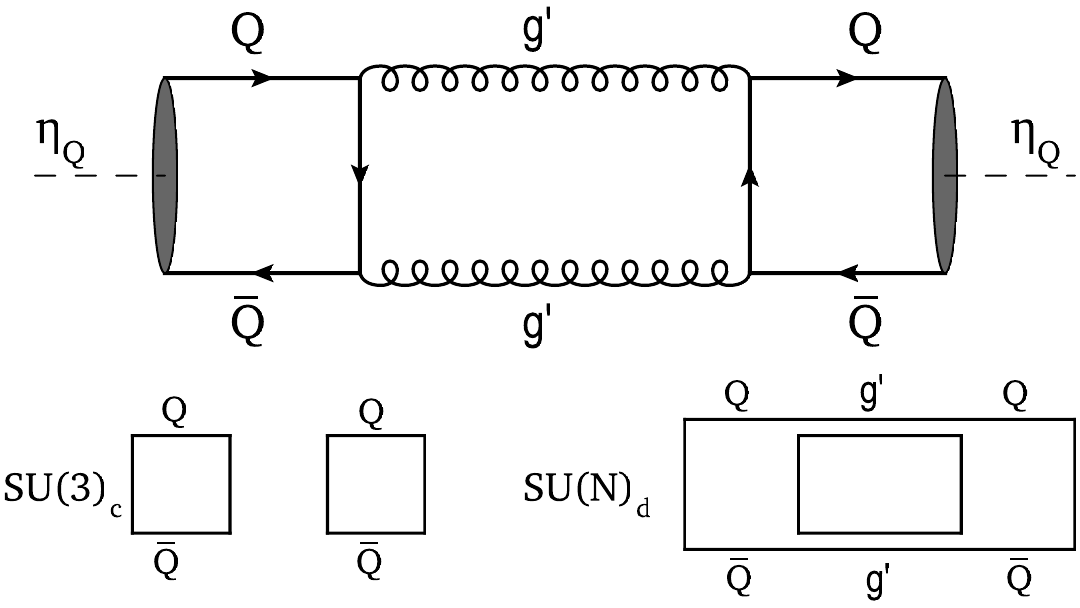}}
\caption{Squared diagram of $\eta_Q\to g'g'$ decay (first line), and the number of closed color loops from contracting the vectorlike quark and dark gluon lines in the large $N_c, N_d$ limit (second line). 
In this process, because both $\eta_Q$ and $g'g'$ are QCD singlets, the QCD color indices carried by $Q, \bar Q$ are closed into two disconnected loops, as indicated by the left part of the second line. On the other hand, the $g'g'$ final states are not necessarily a singlet under the dark $SU(N)_d$. In the large $N_d$ limit, each of the dark gluons can be represented by a double line. As a result, the dark color indices carried by $Q, \bar Q$ can propagate through the $g'$ and they are closed into two loops with a different topology, as shown by the right part of the second line. 
Taking into account of the bound state normalization factor ${1}/{\sqrt{N_c N_d}}$, this decay rate is proportional to $N_c N_d \alpha_d^2$. Similarly, the $\eta_Q\to gg$ decay rate is proportional to $N_c N_d \alpha_s^2$. These power countings agree with our explicit calculation Eq.~(\ref{decays}).
}\label{fig0}
\end{figure}

For large $N_d$, it is much more likely for the $Q, \bar Q$ particles to pair up and form $Q$-onium bound states rather than having $N_d$ of $Q$'s to form baryonic states.~\footnote{There are strong constraints on the relic abundance of dark baryonic states that are stable, carrying electromagnetic charge. See ref.~\cite{Cline:2016nab} and references therein.
To be a singlet under $SU(N)_d$, the $Q$-baryon states must contain $N_d$ of $Q$ particles.
Thus, as a rough estimate, the probability of forming such a baryonic state is $(N_d!)/N_d^{N_d}$ smaller than the probability of forming a $Q\bar Q$ meson state.
For the value of $N_d \gtrsim 100$ of interest to this study, the formation rate of $Q$-baryons in the early universe therefore seems safely small.} 
Here for simplicity, we first consider the case where they all settle down to the $Q$-onium ground states before the annihilation happens.
Because $Q$ is a fermion, there are two ground states, $\eta_Q$ (spin singlet) and $\Upsilon_Q$ (spin triplet), both are singlets under $SU(N)_d$ and $SU(3)_c$.
\footnote{If the $Q\bar Q$ form an octet state under $SU(3)_c$ the strong interaction potential is repulsive and bound state might not form.
On the other hand, the excited $Q$-onium states could de-excite down to $\eta_Q$ by radiating soft dark gluons.}

Because $m_Q\gg \Lambda_d$, $\eta_Q$ and $\Upsilon_Q$ are non-relativistic bound states and their decay rates into gluons in the two sectors are
\begin{eqnarray}\label{decays}
\begin{split}
&\Gamma_{\eta_Q\to gg} = \frac{N_d(N_c^2-1)\alpha_S^2}{4N_cm_Q^2} |R(0)|^2 \ , \\
&\Gamma_{\eta_Q\to g'g'} = \frac{N_c(N_d^2-1)\alpha_d^2}{4N_dm_Q^2} |R(0)|^2 \ , \\
&\Gamma_{\Upsilon_Q\to ggg} = \frac{N_d(N_c^2-1)(N_c^2-4)(\pi^2-9)\alpha_S^3}{36N_c^2 m_Q^2} |R(0)|^2 \ , \\
&\Gamma_{\Upsilon_Q\to g'g'g'} = \frac{N_c(N_d^2-1)(N_d^2-4)(\pi^2-9)\alpha_d^3}{36N_d^2 m_Q^2} |R(0)|^2 \ ,
\end{split}
\end{eqnarray}
where $R(0)$ is the radial part of the bound state wavefunction at the origin, and again $\alpha_S$ and $\alpha_d$ are to be evaluated at the scale $m_Q$.
Because of the generalized Landau-Yang theorem, $\Upsilon_Q$ must decay into three (dark) gluons.
In the decay rates, we show the $N_c \ (=3)$ and $N_d$ dependence on an equal footing.
Here we have neglected the decays of $\eta_Q$ and $\Upsilon_Q$ due to electroweak interactions, assuming they are subdominant.
At leading order, neither $\eta_Q$ nor $\Upsilon_Q$ could decay into $q\bar q$ via a gluon because they are color singlets.

We define a quantity $r$ to be the ratio of entropy production in the visible and dark sectors. 
Clearly, $r$ is proportional to the ratio of decay rates into two sectors.
In the large $N_c$ and $N_d$ limit, for $\eta_Q$ decay, 
\begin{eqnarray}
r_{(\eta)} \equiv \frac{\Gamma_{\eta_Q\to gg}}{\Gamma_{\eta_Q\to g'g'}}\simeq \frac{\alpha_S^2}{\alpha_d^2} \ ,
\end{eqnarray}
with all the color pre-factors cancelled out; and for $\Upsilon_Q$ decay, 
\begin{eqnarray}
r_{(\Upsilon)} \equiv  \frac{\Gamma_{\Upsilon_Q\to ggg}}{\Gamma_{\Upsilon_Q\to g'g'g'}} \simeq \frac{N_c\alpha_S^3}{N_d \alpha_d^3} \ .
\end{eqnarray}
Because the gauge couplings satisfy $\alpha_S \sim 1/N_c$ and $\alpha_d \sim 1/N_d$, we find that $r_{(\eta)} \sim r_{(\Upsilon)}$.
The actual value of $r$ should lie between $r_{(\eta)}$ and $r_{(\Upsilon)}$.
This is still correct when the excited $Q\bar Q$ states are taken into account because they must either de-excite down to the ground states or directly annihilate decay into two or three (dark) gluons.

The above large $N_d$ power counting is in strong contrast with that in Eq.~(\ref{eq5}) where all the $Q\bar Q$ annihilation cross sections are of the same order. The key here is that after the dark $SU(N)_d$ confinement, $Q, \bar Q$ must annihilate in the color singlet state, and this greatly reduces the color degrees of freedom of the final state dark gluons from $N_d^2$ to $1$. As a consequence, the annihilating decay into the dark sector is suppressed.

It is important to notice that $r$ is independent of the vectorlike quark mass $m_Q$.
The ratio of energy dumped into the two sectors is dictated by the ratio of decay rates, which leads to a ``selection rule'', $r \sim N_d^2$. 
Counter-intuitively, the dark sector gets less reheated when it contains more degrees of freedom in the presence of a large $N_d$.
Such a power counting for the decays of $\eta_Q$ (and similarly for $\Upsilon_Q$) can also be understood from the number of closed color loops in the square of the decay amplitude, as depicted in Fig.~\ref{fig0}.

The decay lifetimes of the $Q$-onium states are dictated by the mass scale of $Q$ and are much shorter than the cosmological time scale.
In the instantaneous confinement and decay approximation, the energy densities of the two sectors are reheated to
\begin{equation}\label{reheat}
\begin{split}
\hat\rho_{\rm d} (a_c) &= \rho_{\rm d}(a_c) + \rho_{Q+\bar Q} (a_c) \frac{1}{1+r} \ ,\\
\hat\rho_{\rm SM} (a_c) &= \rho_{\rm SM}(a_c) + \rho_{Q+\bar Q} (a_c) \frac{r}{1+r} \ .
\end{split}
\end{equation}
With these, we can also solve for the temperatures in the two sectors after the reheating, $\mathcal{T}_r$ and $T_r$, respectively,
\begin{equation}\label{Tr}
\mathcal{T}_r = \left[\frac{30\hat\rho_{\rm d} (a_c)}{2\pi^2 (N_d^2-1)}\right]^{\frac{1}{4}} \ ,\ \ \
T_r = \left[\frac{30\hat\rho_{\rm d} (a_c)}{\pi^2 g_*^{\rm SM}(T_r)}\right]^{\frac{1}{4}} \ .
\end{equation}

\begin{figure*}[t]
\centerline{\hspace{0.5cm}\includegraphics[width=2.1\columnwidth]{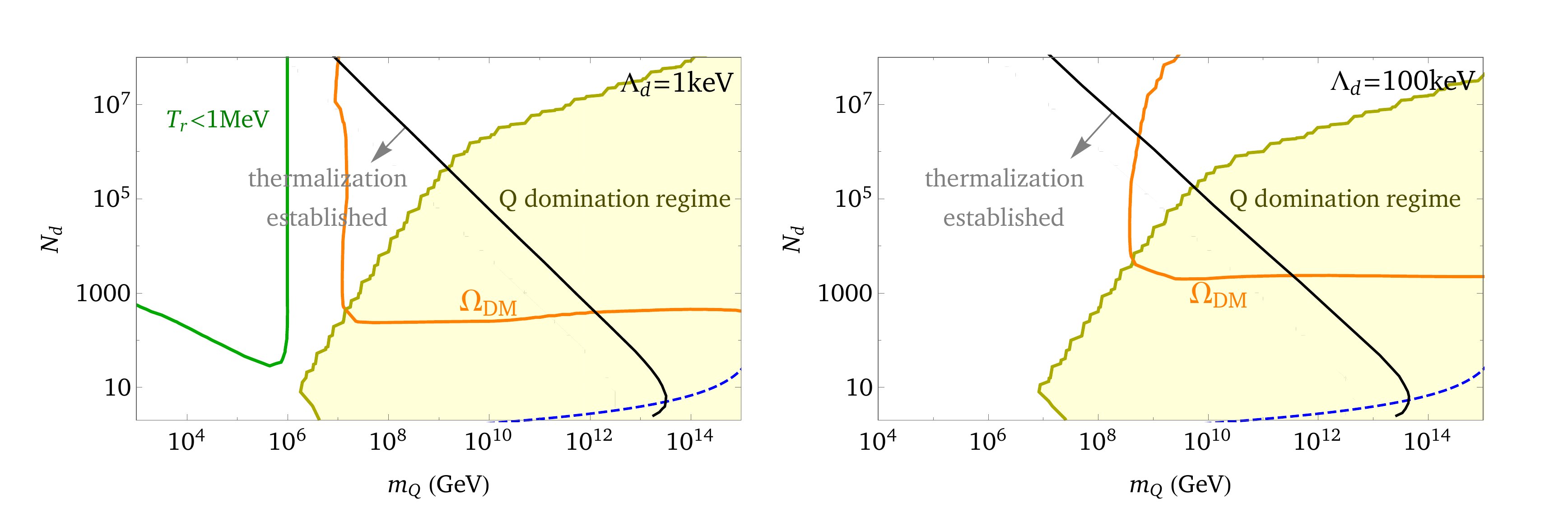}}
\caption{In the parameter space of $N_d$ versus $m_Q$, the orange curve in the two plots above corresponds to the correct relic abundance (see Eq.~(\ref{obs})) for the $SU(N)_d$ glueball DM, which becomes independent of the vectorlike quark mass $m_Q$ when they are heavy enough (the horizontal part of the curve). Such UV insensitiveness occurs as long as there is a stage in the early universe when the energy density of the vectorlike quarks becomes dominant, as shown by the yellow shaded regions. The black curve is a boundary below which the vectorlike quarks and the DM and dark sector can reach thermal equilibrium at temperature around $m_Q$. 
In the hatched region above the black curve, we make the assumption that the vectorlike quarks are abundant enough so that their domination in the universe is guaranteed to happen. We have fixed the value of the dark sector intrinsic scale $\Lambda_d=1\, (100)$\,keV in the left (right) plot. In the left plot, the region surrounded by the green curve has too low reheating temperature after the annihilation decay of the vectorlike quarks. In both plots, along the blue curve, one may realize unification of SM gauge couplings in the presence of the vectorlike quarks.
}\label{fig1}
\end{figure*}

After the $Q$-onium states all decayed away, both $\rho_{\rm d}$ and $\rho_{\rm SM}$ will redshift as radiation, $\sim a^{-4}$.
As the universe further expands, a second phase transition will occur in the dark sector when the temperature there equals the intrinsic scale of $SU(N)_d$, {\it i.e.}, $\mathcal{T} =\Lambda_d$.
The dark gluons will then confine into glueballs.~\footnote{Because the $Q$ number density is Boltzmann suppressed after freeze out and the dark gluons are reheated by the $Q$-onium decay, the confinement of dark gluons always happens later than the confinement of $Q, \bar Q$.}
The lightest dark glueball state, $\phi$, is our DM candidate, with mass $m_\phi \sim \Lambda_d$. 
The expansion parameter associated with the dark glueball formation will be called, $a_\Lambda$, which is equal to
\begin{eqnarray}
a_\Lambda = a_c \left( \frac{15 \hat \rho_{\rm d}(a_c)}{\pi^2 (N_d^2-1) \Lambda_d^4} \right)^{\frac{1}{4}} \ .
\end{eqnarray}
The corresponding SM sector temperature $T_\Lambda$ can be solved using, $g_{*S}^{\rm SM}(T_r) T_r^3 a_c^3 = g_{*S}^{\rm SM}(T_\Lambda) T_\Lambda^3 a_\Lambda^3$.

We assume the dark glueball energy density simply matches to that of the dark gluons,
\begin{eqnarray}
\rho_{\phi}(a_\Lambda) = \rho_{\rm d}(a_\Lambda) = \frac{\pi^2}{15} (N_d^2-1) \Lambda_d^4 \ ,
\end{eqnarray}
and after this phase transition in the dark sector, the dark glueball quickly turns non-relativistic and its number density redshifts like pressure-less matter,~\footnote{For dark sector temperatures below $\Lambda_d$, it is possible for the glueball DM to further slightly decrease their comoving number through the $3\phi\to2\phi$ scattering, until such a process freezes out. The conserved quantity is $\phi$'s entropy. Because $\phi$ is already non-relativistic at this stage, the net reduction of $\phi$ particles is small~\cite{Carlson:1992fn, Soni:2016gzf}. In our calculation, we neglect this effect.} until today. Therefore,
\begin{eqnarray}\label{rhoD}
\rho_{\phi}(a_0) \simeq \rho_{\phi}(a_\Lambda) \left( \frac{a_\Lambda}{a_0} \right)^3 \ ,
\end{eqnarray}
where $a_0$ stands for today. On the other hand, the energy density in the SM sector remains radiation-like until today. Again, taking into account of the possible change in the number of degrees of freedom, we get
\begin{eqnarray}\label{rhoSM}
\rho_{\rm SM}(a_0) = \rho_{\rm SM}(a_\Lambda) \left( \frac{g_*(T_0)}{g_*(T_\Lambda)} \frac{g_{*S}(T_\Lambda)}{g_{*S}(T_0)}  \right) \left( \frac{T_0}{T_\Lambda} \right) 
\left( \frac{a_\Lambda}{a_0} \right)^3 \ , \nonumber \\
\end{eqnarray}
where $T_0=2.7\,$K is the CMB photon temperature.

The ratio of the DM energy density to that in the SM sector calculated from Eqs.~(\ref{rhoD}) and (\ref{rhoSM}) is to be compared with 
the observed value today~\cite{pdg}
\begin{eqnarray}\label{obs}
\frac{\rho_\phi}{\rho_{\rm SM}} = \left.\frac{\rho_\phi}{1.68\rho_{\gamma}}\right|_{obs.} \simeq 3\times 10^3 \ ,
\end{eqnarray}
where we have assumed the standard ratio of neutrino to photo temperature, $T_\nu = (4/11)^{1/3} T_\gamma$. 

\section{A Cosmic Selection Rule}

Our results are shown in Fig.~\ref{fig1} in the parameter space of $N_d$ versus $m_Q$, for two fixed values of $\Lambda_d=1, 100\,$keV, respectively.
Along the orange curves, the glueball DM in this model could obtain the correct relic abundance. Below and to the left of the orange curves, the glueball DM is overproduced.

Based on these results, we find an interesting parameter space for the correct dark glueball relic abundance, corresponding to the horizontal part of the orange curves, where the vectorlike quarks are heavy enough and abundant enough.
In this case, the energy density of the vectorlike quarks came into domination of the universe for a while before they hadronize and annihilate decay away (corresponds to the yellow shaded region in Fig.~\ref{fig1}). A quite intriguing feature here is that, the relic abundance depends very weakly on $m_Q$. Instead, it is only determined by the intrinsic dark scale $\Lambda_d$ and the number of dark colors $N_d$.

\begin{figure}[t]
\centerline{\includegraphics[width=1\columnwidth]{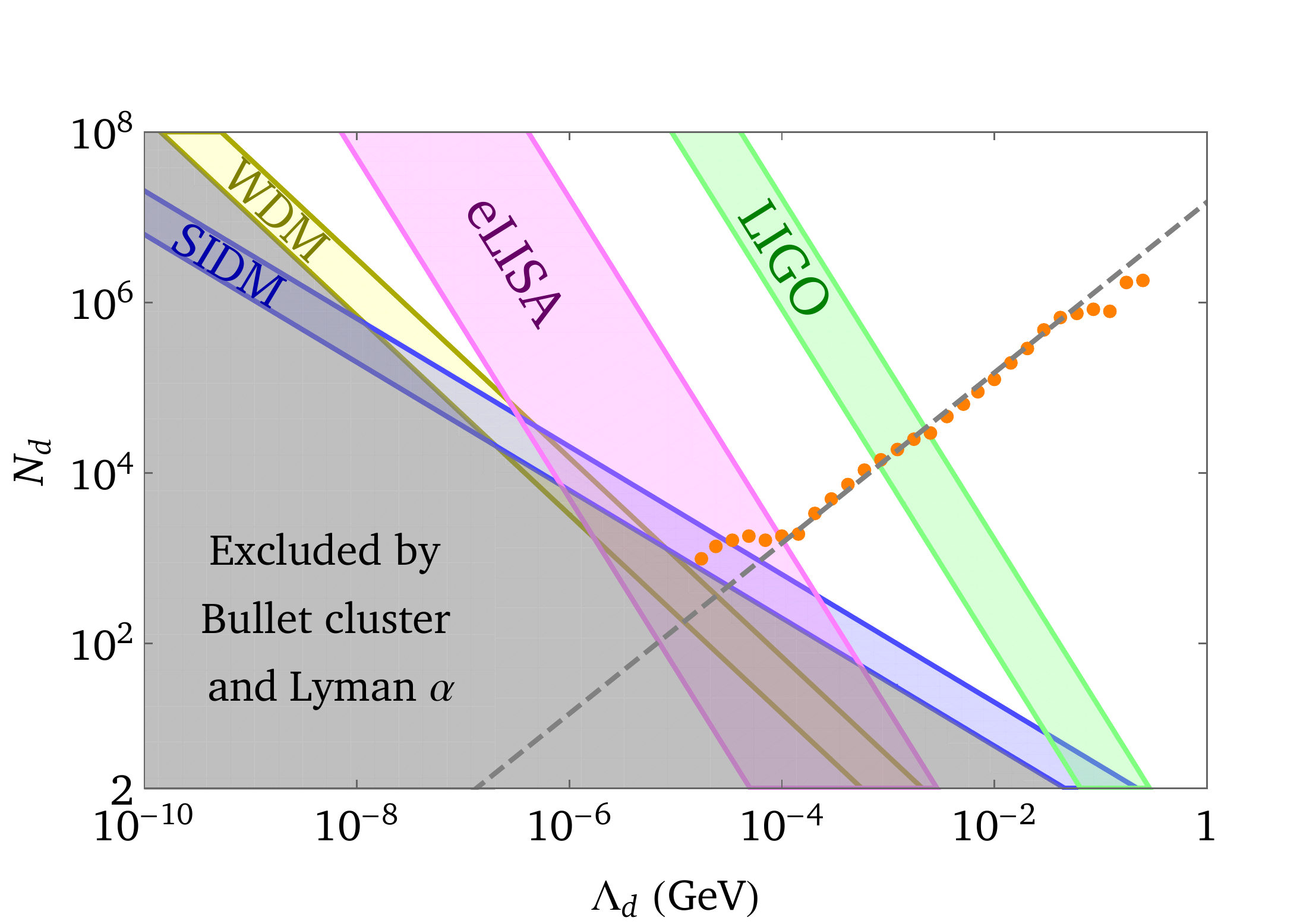}}
\caption{The low energy parameter space of $N_d$ versus $\Lambda_d$. The correct relic abundance to the glueball DM can be accommodated on the orange dots, where we assumed the vectorlike quarks to be heavy and abundant enough to dominate the early universe. The black dashed line corresponds to the approximate relation derived in Eq.~(\ref{selection}), as a fit to the orange dots.
Also shown are the regions of interest to astrophysical probes, including warm DM, self-interacting DM, as well as possible gravitational wave signals from massive compact objects formed by the glueball condensate, as derived in refs.~\cite{Soni:2016gzf} and \cite{Soni:2016yes}.
The gray shaded region in the lower-left corner (below the yellow or blue band) is excluded by astrophysical observations. 
}\label{fig2}
\end{figure}

The key reason for this feature follows from observing Eq.~(\ref{reheat}). We consider the limit where the second term on the right-hand side dominates in both energy densities, {\it i.e.}, $\hat\rho_{\rm SM} (a_c) \simeq r \hat\rho_{\rm d} (a_c)$.
In this limit, up to changes in $g_*^{\rm SM}$, the ratio of Eqs.~(\ref{rhoD}) and (\ref{rhoSM}) is given by,
\begin{eqnarray}\label{ratio}
\frac{\rho_{\phi}(a_0)}{\rho_{\rm SM}(a_0)} \sim \frac{T_\Lambda}{r T_0} \ .%\sim \frac{\Lambda_d}{T_0 N [g^{\rm SM}_*(T_\Lambda)]^{1/4}} \ .
\end{eqnarray}
It is worth noting again that the right-hand side does not depend on the vectorlike quark mass $m_Q$.
The value of $r$ is determined by Eq.~(\ref{decays}), which goes as $\sim N_d^2$ in the large $N_d$ limit.
The value of $T_\Lambda$, in unit of $\Lambda_d$, can be inferred from the relation between $\mathcal{T}_r$ and $T_r$ calculated using Eq.~(\ref{Tr}).
Again, up to changes in $g_*^{\rm SM}$, we find
\begin{eqnarray}\label{TLambda}
T_\Lambda \sim \Lambda_d \left( \frac{r N_d^2}{g_*^{\rm SM}} \right)^\frac{1}{4} \simeq \frac{N_d\Lambda_d}{(g_*^{\rm SM})^\frac{1}{4}} \ .
\end{eqnarray}
Based on Eqs.~(\ref{obs}), (\ref{ratio}) and (\ref{TLambda}), we find a correlation between $N_d$ and $\Lambda_d$ that yields the correct DM relic abundance, 
\begin{eqnarray}\label{selection}
N_d \simeq c\,\Lambda_d \ ,
\end{eqnarray}
where we find numerically, $c\simeq 15\,{\rm keV}^{-1}$.
The relation between $N_d$ and $\Lambda_d$ is depicted in Fig.~\ref{fig2}.
For small values of $\Lambda_d$, the required $N_d$ for relic density is slightly larger than the prediction in Eq.~(\ref{selection}). This is mainly because
the temperature at $T_\Lambda$ is below the QCD phase transition, where annihilating away the hadronic sector tends to make the rest of SM particle ``hotter'' than estimated above. As a result, a larger value of $N_d$ is needed to keep the ratio Eq.~(\ref{ratio}) invariant.

It is worth reminding our reader that the dark $SU(N)_d$ intrinsic scale $\Lambda_d$ is closely tied to the the glueball DM mass. 
As the main finding of this work, for given DM mass in this model there is a particular value of $N_d$ for its relic abundance today,
{\it i.e.}, the branching ratios of the vectorlike quarkonium decay in the early universe into the SM and dark sectors only depend on $N_d$ but are
largely independent of the high scale parameters such as the vectorlike quark mass.
Therefore, we would like to call it {\it a cosmic selection rule} for the $SU(N)_d$ glueball dark matter.

We have taken into account a consistency condition which states that the decay of $Q$-onia must reheat the SM sector to a temperature above MeV scale so that the standard big-bang nucleosynthesis could begin. This is more relevant for lower values of $\Lambda_d$ (in the left plot of Fig.~\ref{fig1}, the region enclosed by the green curve is excluded), where the vectorlike quarks could be diluted for a longer period before they decay.

Another feature of the above cosmic selection rule is that the desired value of $N_d$ increases with $\Lambda_d$. 
On the other hand, for $T\gg m_Q$, the interaction strength between the two sectors are more suppressed at larger $N_d$ (see Eq.~(\ref{eq5})).
As a result, the available parameter space for the above mechanism to work with a thermal initial condition shrinks for larger $\Lambda_d$.
This feature can be seen in Fig.~\ref{fig1}, where the overlap between the orange curve (for correct relic density) and the yellow region (for $Q$ domination) shrinks for larger $\Lambda_d$, unless one gives up the thermal initial condition (see next paragraph).
If one insists on the thermal initial condition, there is an upper bound on $\Lambda_d$ for our mechanism to work, which is $\Lambda_d\lesssim 0.2$\,GeV. This is why the range of the orange dots is limited from the right in Fig.~\ref{fig2}.

Finally, we want to point out that our findings can be applied to more general initial conditions of the early universe. 
Our discussions so far begin with the vectorlike quarks, SM and dark sectors all thermalized in together.
The key ingredient here is to have the vectorlike quarks to dominate the energy density of the early universe.
It could also be realized with non-thermal initial conditions. As long as the vectorlike quarks are abundantly produced, the same conclusion will remain. In Fig.~\ref{fig1}, we have extended the relic abundance calculation into the hatched region above the black curve, which shows the same cosmic selection rule still holds.

\section{Implications of Our Results}

%In the past two sections, we derive a correlation between the dark $SU(N)_d$ scale, $\Lambda_d$, and the number of dark colors, $N_d$, for obtaining the correct relic density for glueball DM. This result depends on the story at the heavy vectorlike quark mass scale, but does not depends on the p
%

The hidden glueball DM is based on very simple assumptions about the dark sector, but could leave quite a few imprints in the cosmology and astrophysics.
They include the self-interacting and warm DM scenarios as the consequences of the scalar glueball potential interactions, as well as the possibility of 
forming macroscopic compact objects (dark stars) which may source the gravitational waves.
Fig.~\ref{fig2} also shows the regions in the $N_d$ versus $\Lambda_d$ parameter space that could accommodate these phenomena~\cite{Soni:2016gzf, Soni:2016yes}. 
These possibilities serve as incentives for future experimental tests.
In contrast, in the cosmic selection rule Eq.~(\ref{selection}), the relationship between $N_d$ and $\Lambda_d$ is derived from a different perspective by requiring
a UV insensitive glueball relic density. It could serve as a theoretical guide on the favored model parameter space.

A potential consequence of the $Q, \bar Q$ matter domination in the early universe is the formation of minihalos and even primordial black holes, because the primordial density perturbations grow linearly during that period (see~\cite{scott, Zhang:2015era} for discussions along this direction but within different contexts).
The mass and size of these objects are typically the size of the universe around the reheating temperature after the $Q$-onia decay.

As the final remark, we discuss the impact of our result on the running of gauge couplings. With a large dark color number $N_d$, the running of dark $SU(N)_d$ coupling is slow but still asymptotic free. On the other hand, it will have a strong impact in the running of SM gauge couplings at energy scale above the vectorlike quark mass. 
The electroweak quantum numbers of the vectorlike quarks chosen in Eq.~(\ref{quantumnumbers}) are not important for the cosmological discussions above,
but as we show below, such a particular choice allows the SM gauge couplings to unify better at much higher scales.
The vectorlike quark contributions to the beta equations at one loop level are
\begin{eqnarray}
\Delta b_1= \frac{2}{15}N_d, \ \ \ \Delta b_2=2N_d, \ \ \ \Delta b_3=\frac{4}{3}N_d \ ,
\end{eqnarray} 
where the $b_i$'s are defined as, ${d \alpha_i}/({d\ln\mu})= b_i \alpha_i^2/(2\pi)$.
In fact, we find that with $Q$ and $\bar Q$ it is possible to obtain unification of the three couplings if $N_d\simeq 21.4/\left(35.4-\ln [m_Q/{\rm GeV}]\right)$. This relation between $N_d$ and $m_Q$ is shown by the blue dashed curves in Fig.~\ref{fig1}. The unification scale associated with this curve is around $2.3\times10^{15}\,$GeV, which is close to the present proton decay bound (see, {\it e.g.},~\cite{Harigaya:2016vda} and references therein), given the uncertainties of GUT scale threshold corrections to the scale. To obtain the correct glueball relic abundance, we find the needed value of $N_d$ turns out to be much higher than that for unification. As a result, at energy scales above $2m_Q$, all the SM gauge couplings would run quickly into the non-perturbative regime. This implies that the simple model we have discussed above is only effective at describing the low energy physics perturbatively, but not at energy scales and temperatures above the $2m_Q$ scale. We do not aim at deriving a high scale theory here, but rather stress again that our cosmological calculation in this work only involves physics below the $m_Q$ scale and is largely insensitive to the UV physics, and moreover, the final glueball DM relic density almost does not depend on $m_Q$ either.

\section{Acknowledgement}

The work of AS is supported in part by the DOE Grant No. DE-AC-02-98CH10886.
The work of YZ is supported by the DOE grant DE-SC0010143.
YZ would like to thank A. Pierce and Y. Zhao for useful discussions, and James Cline for bringing up the constraints on stable electric-charged exotic particles and ref.~\cite{Cline:2016nab}.

\end{document}